# Power-dependent shaping of vortex solitons in optical lattices with spatially modulated nonlinear refractive index

Yaroslav V. Kartashov, Victor A. Vysloukh, and Lluis Torner

*ICFO-Institut de Ciencies Fotoniques, and Universitat Politecnica de Catalunya, Mediterranean Technology Park, 08860 Castelldefels (Barcelona), Spain*

We address vortex solitons supported by optical lattices featuring modulation of both the linear and nonlinear refractive indices. We find that when the modulation is out-of-phase the competition between both effects results in remarkable shape transformations of the solitons which profoundly affect their properties and stability. Nonlinear refractive index modulation is found to impose restrictions on the maximal power of off-site solitons, which are shown to be stable only below a maximum nonlinearity modulation depth.

Vortex solitons are excited states carrying nested phase singularities. In a uniform focusing medium, such excited states are highly prone to azimuthal instabilities [1]. Vortices are stabilized in discrete systems [2,3] and in continuous periodic lattices [4-6]. Other types of non-periodic optically induced refractive index landscapes may also support stable vortex solitons [7,8]. Up to now, vortex lattice solitons were considered mostly in structures whose nonlinearity is uniform in the transverse plane. However, current technologies allow fabrication of lattices with simultaneous modulation of linear and nonlinear refractive indices. Among the examples is a two-dimensional array written in glass by high intensity fs laser pulses, where optical damage results in an increase of the linear refractive index accompanied by remarkable decrease of nonlinear refractive index [9]. Such structures with out-of-phase modulation of linear and nonlinear refractive indices are especially interesting because the competition between linear and nonlinear effects is controlled by the power of the input beams. One-dimensional solitons in lattices with nonuniform nonlinearity have been extensively studied in both discrete [10,11] and continuous [12-18] settings. Recently, two-dimensional solitons were analyzed in purely nonlinear lattices [19] and in discrete system with nonlinear coupling [20].

In this Letter we address the properties and stability of vortex solitons in continuous two-dimensional lattices with an out-of-phase modulation of the linear and nonlinear refractive indices. We uncover that vortex solitons in such lattices undergo unusual shape transformations at high powers. Also, they are found to be stable only below a maximum nonlinearity modulation depth.

We use the nonlinear Schrödinger equation to describe the propagation of a laser beam along the $\xi$ axis of a two-dimensional lattice with an out-of-phase modulation of both, the linear and nonlinear refractive indices. Namely,

$$i\frac{\partial q}{\partial \xi} = -\frac{1}{2}\left(\frac{\partial^2 q}{\partial \eta^2} + \frac{\partial^2 q}{\partial \zeta^2}\right) - [1 - \sigma R(\eta,\zeta)]|q|^2 q - pR(\eta,\zeta)q. \quad (1)$$

Here $q$ is the dimensionless amplitude of the light beam; $\eta, \zeta$ and $\xi$ are the transverse and longitudinal coordinates, respectively; $p$ and $\sigma$ are the depths of modulation of linear and nonlinear refractive indices, while the function $R(\eta,\zeta) = \sin^2(\Omega\eta)\sin^2(\Omega\zeta)$ describes the lattice shape. The nonlinear coefficient $1 - \sigma R$ attains its minima at the points where the refractive index has a maximum.

We search for vortex soliton solutions of Eq. (1) characterized by their conserving power $U = \int\int_{-\infty}^{\infty} |q|^2 \, d\eta d\zeta$ in the form $q = (u + iv)\exp(ib\xi)$, where $u$ and $v$ are the real and imaginary parts, respectively, $b$ is the propagation constant. In this Letter we focus on so-called off-site vortices [2-6] with topological charge, or winding number, $m = 1$. We set $p = 8$, $\Omega = 2$, and vary the nonlinearity modulation depth $\sigma$. Substitution of perturbed soliton solutions in the form $q = [u + iv + u_p \exp(\delta\xi) + iv_p \exp(\delta\xi)]\exp(ib\xi)$, where $u_p(\eta,\zeta), v_p(\eta,\zeta)$ are small perturbations, into Eq. (1) and linearization yields the eigenvalue problem

$$\delta u_p = -\frac{1}{2}\left(\frac{\partial^2 v_p}{\partial \eta^2} + \frac{\partial^2 v_p}{\partial \zeta^2}\right) + bv_p - (1 - \sigma R)v_p(u^2 + 3v^2) - 2(1 - \sigma R)uvu_p - pRv_p,$$
$$\delta v_p = \frac{1}{2}\left(\frac{\partial^2 u_p}{\partial \eta^2} + \frac{\partial^2 u_p}{\partial \zeta^2}\right) - bu_p + (1 - \sigma R)u_p(3u^2 + v^2) + 2(1 - \sigma R)uvv_p + pRu_p, \quad (2)$$

that we solved numerically to find the growth rate $\delta = \delta_r + i\delta_i$.

Different shapes of the obtained off-site vortex solitons are shown in Fig. 1. At moderate power levels, when effects of nonlinear and linear refraction are comparable, the field modulus distribution features four bright spots whose positions coincide with the local maxima of the linear lattice [Fig. 1(a)]. The phase distribution is spiral-staircase-like and the phase changes by $2\pi$ around any closed contour surrounding the singularity located at $\eta, \zeta = 0$. Vortex solitons exist above a power threshold and above the corresponding cutoff on propagation constant $b_{\mathrm{co}}$; their power is a non-monotonic function of $b$ [Fig. 2(a)]. If the lattice depth $p$ increases, the cutoff rapidly grows, while the minimal power drops off. Interestingly, a growth of the nonlinearity modulation depth $\sigma$ does affect the minimal power, which increases with $\sigma$, despite the fact that the amplitude of vortex solitons close to cutoff is small and that light expands substantially across the lattice.

Power growth results in progressive localization of bright spots forming the soliton. If the nonlinearity modulation depth is small enough (i.e., $\sigma \lesssim 0.5$ at $p = 8$) the bright spots always stay in the vicinity of the linear lattice maxima. However, at $\sigma \gtrsim 0.5$ the competition between linear and nonlinear refraction might result in concentration of light in the regions where nonlinearity is stronger [Fig. 1(b)]. Even in this case the vortex solitons conserve their staircase-like phase distribution. Such shape transformation is accompanied by a decrease of the soliton power for $b$ values above a critical value $b_{\mathrm{cr}}$ [Fig. 2(a)]. Thus, nonlinearity modulation imposes a restriction on the maximal power of off-site vortex solitons. In the interval $0.5 \lesssim \sigma \lesssim 0.9$ the maximal power slightly increases with $\sigma$. Note that $b_{\mathrm{cr}}$ drops off rapidly with $\sigma$ [Fig. 2(c)] and for $\sigma \gtrsim 1.3$ the vortex power starts diminishing monotonically with $b$ in the entire existence domain. Nevertheless, one should keep in mind that the values of $\sigma > 1$ correspond to transition to defocusing nonlinearity in the centers of lattice channels, something that does not occur in currently available optical lattices.

For $\sigma \gtrsim 0.9$ the bright vortex spots tend to fuse into modulated ring-like structures for $b > b_{\mathrm{cr}}$, i.e. high-power vortices tend to contract [Fig. 1(c)] rather than to expand as it happens for smaller $\sigma$ [Fig. 1(b)]. In fact, for $\sigma \gtrsim 0.9$ one can easily identify an upper vortex soliton branch that looks like an extrapolation of the lower branch to higher powers [Fig. 2(b)]. The amplitude profiles of vortex solitons from the upper and lower branches almost coincide at $b = b_{\text{cr}}$, but with increase of $b$ the corresponding solitons acquire substantially different shapes because of the expansion of solitons from the upper branch and the contraction of solitons from the lower branch [compare, e.g., Figs. 1(c) and 1(d)].

The modulation on nonlinear refractive index profoundly affects the stability of vortex solitons. As in the case of linear lattices, vortex solitons feature a rather narrow domain of oscillatory instability close to their cutoff [Fig. 2(e)]. However, there appears another one, which is linked with vortex shape transformations. The perturbation growth rate $\delta_r$ might become positive already for $b < b_{\text{cr}}$, i.e. below the point where $dU/db$ changes the sign [Fig. 2(f)]. At $b = b_{\text{cr}}$ the slope of the $\delta_r(b)$ dependence grows remarkably and the exponential instability replaces the oscillatory one. Therefore, vortex solitons in nonlinear lattices are stable only within the certain interval of propagation constants $b_{\text{low}} \leq b \leq b_{\text{upp}}$ [Fig. 2(d)]. Since $b_{\text{low}}$ increases with $\sigma$ and $b_{\text{upp}}$ decreases with $\sigma$, the stability domain shrinks completely at $\sigma \gtrsim 0.79$. Thus, off-site vortex solitons can be stable in nonlinear lattices only when the nonlinearity modulation depth does not exceed the critical value. This critical value grows with increase of the linear lattice depth $p$. Solitons from the upper branch are always unstable [Fig. 2(b)].

Direct integration of Eq. (1) confirms the results of linear stability analysis. Low-power vortices with $b < b_{\text{low}}$ decay via progressively growing oscillations of their bright spots amplitudes and they usually transform into fundamental solitons [Fig. 3(a)]. Vortices whose propagation constants are located within the stability region maintain their shapes and energy circulation over indefinitely long distance even when they are strongly perturbed [Fig. 3(b)]. Finally, high-power vortex solitons with $b > b_{\text{upp}}$ suffer from a specific instability: While perturbed with input noise their spots tend to shift from maxima of the linear lattice into the regions where nonlinearity is stronger (drift instability), where they experience rapid collapse after just a few diffraction lengths [Fig. 3(c)].

Thus, the central result of this Letter is that out-of-phase modulation of the linear and nonlinear refractive indices gives rise to important, new power-dependent shape transformations of vortex lattice solitons. Such solitons exist and can be stable only within a limited range of light intensities. Also, they are found to be stable only below a maximum nonlinearity modulation depth.

# References with titles

# References without titles

# Figure captions

Figure 1. Field modulus distributions for vortex solitons at (a) $b=5$, $\sigma=0.7$, (b) $b=9.6$, $\sigma=0.7$, (c) $b=6$, $\sigma=1$, lower branch, and (d) $b=6$, $\sigma=1$, upper branch. White lines indicate positions of maxima of linear lattice.

Figure 2. Power versus $b$ at (a) $\sigma=0.5$ and $0.7$, and (b) $\sigma=1$. Circles in (a) correspond to solitons in Figs. 1(a) and 1(b), while circles in (b) correspond to solitons from Figs. 1(c) and 1(d). (c) Cutoff and critical propagation constant versus $\sigma$. (d) Stability domain (shaded) for vortex solitons on $(\sigma,b)$ plane. $\delta_r$ versus $b$ for $b$ values close to cutoff (e) and $b$ values close to inflection point of $U(b)$ curve (f).

Figure 3. Propagation dynamics of perturbed vortex solitons with (a) $b=3.2$, (b) $5$, and (c) $7$ at $\sigma=0.7$. White noise with variance $\sigma^2_{\text{noise}}=0.01$ was added into input distributions. Field modulus distributions are shown at different distances.

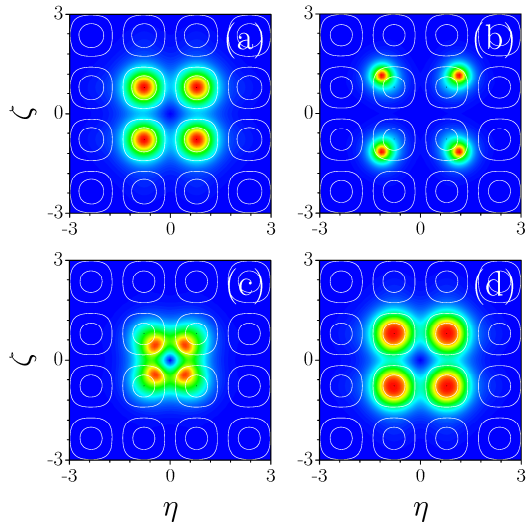

Figure 1. Field modulus distributions for vortex solitons at (a) $b=5$, $\sigma=0.7$, (b) $b=9.6$, $\sigma=0.7$, (c) $b=6$, $\sigma=1$, lower branch, and (d) $b=6$, $\sigma=1$, upper branch. White lines indicate positions of maxima of linear lattice.

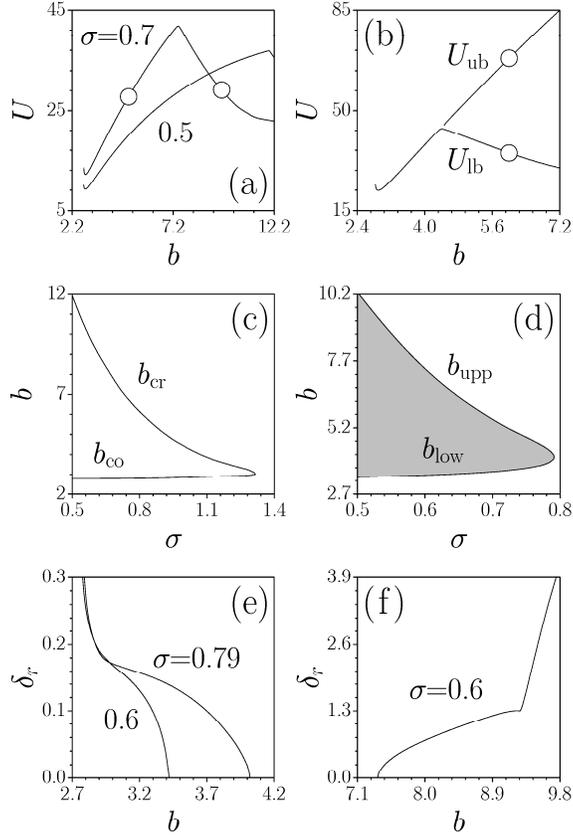

Figure 2. Power versus $b$ at (a) $\sigma = 0.5$ and $0.7$, and (b) $\sigma = 1$. Circles in (a) correspond to solitons in Figs. 1(a) and 1(b), while circles in (b) correspond to solitons from Figs. 1(c) and 1(d). (c) Cutoff and critical propagation constant versus $\sigma$. (d) Stability domain (shaded) for vortex solitons on $(\sigma, b)$ plane. $\delta_r$ versus $b$ for $b$ values close to cutoff (e) and $b$ values close to inflection point of $U(b)$ curve (f).

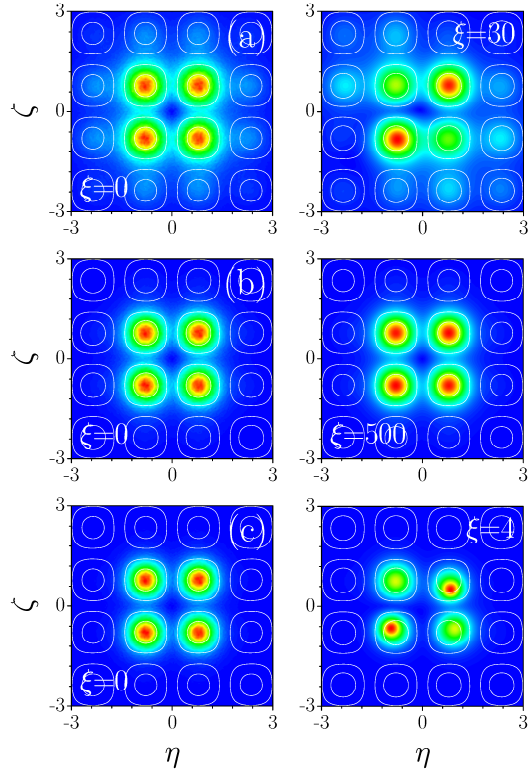

Figure 3. Propagation dynamics of perturbed vortex solitons with (a) $b = 3.2$, (b) $5$, and (c) $7$ at $\sigma = 0.7$. White noise with variance $\sigma_{\text{noise}}^2 = 0.01$ was added into input distributions. Field modulus distributions are shown at different distances.

11